\documentclass[a4paper,10pt,twoside]{cpc-hepnp}
\usepackage{multicol}
\usepackage{multirow}
\usepackage{graphicx}
\usepackage{booktabs}
\usepackage{amssymb,bm,mathrsfs,bbm,amscd}
\usepackage[tbtags]{amsmath}
\usepackage{lastpage}
\usepackage{CJK}

\begin{document}
\begin{CJK*}{GBK}{song}

\title{Preliminary lattice study of  $I=0$ $K \overline{K}$ scattering
\thanks{
Supported by Fundamental Research Funds for the Central Universities (2010SCU23002).
}
}

\author{
Ziwen Fu$^{\dag}$
}
\maketitle
\footnotetext{\hspace*{-0.2cm}\footnotesize
$^\dag$ Corresponding author, E-mail: fuziwen@scu.edu.cn
}
\address{
Key Laboratory of Radiation Physics and Technology {\rm (Sichuan University)},
Ministry of Education;
Institute of Nuclear Science and Technology, College of Physical Science and Technology,
Sichuan University,
Chengdu 610064, P. R. China.
}

\begin{abstract}
We deliver the realistic ab initio lattice investigations of
$K \overline{K}$  scattering.
In the Asqtad-improved staggered dynamical fermion formulation,
we carefully measure $K\overline{K}$ four-point function in the $I=0$ channel
by moving wall sources without gauge fixing,
and clearly find an attractive interaction in this channel,
which is in agreement with the theoretical predictions.
An essential ingredient in our lattice calculation is to properly
treat the disconnected diagram. Moreover, we explain the difficulties of these lattice calculations, and discuss the way to improve the statistics.
Our lattice investigations are carried out with the MILC $2+1$ gauge configuration
at lattice spacing $a \approx 0.15$~fm.
\end{abstract}

\begin{pacs}
{12.38.Gc, 13.75.Lb, 11.15.Ha}
\end{pacs}


\begin{multicols}{2}

\section{Introduction}
The $K \overline{K}$ interactions are hitherto poorly understood~\cite{Fu:2012ng,Doring:2011vk},
our principal incentive  for lattice study of $K \overline{K}$ scattering
was preliminarily explained in Ref.~\cite{Fu:2012ng},
as an effort to reveal the perplexing properties of the scalar resonances
$a_0(980)$ and $f_0(980)$,
whose masses are in the vicinity of $K \overline{K}$ threshold.
Some experiments of the $K \overline{K}$ scattering
were carried out at the cooler synchrotron COSY
facilities~\cite{Winter:2006vd,Dzyuba:2008fi,Maeda:2008mx,
Silarski:2009yt,Silarski:2010ae,Xie:2010md},
which suggests a vigorous clue of $K \overline{K}$ background.
Moreover, the $K\overline{K}$ interactions are believed to be
fundamental process to many intriguing physical phenomena, for instance,
kaon condensation in neutron stars~\cite{Li:1997zb},
and a lattice investigation on
kaon condensation was explored~\cite{Detmold:2008yn}.

Though the study of $K \overline{K}$ scattering from QCD
is basically a non-perturbative problem,
some theoretical (or phenomenological) attempts are being made
to examine $K\overline{K}$
scattering~\cite{Oller:1998hw,Guerrero:1998ei,Guo:2011pa,Kaminski:2009qg,Xiao:2001pt,Jia:2011zzd}.
Lattice QCD is currently the unparalleled non-perturbative method which
enables us to learn $K\overline{K}$ scattering from first principles.
Moreover, due to  the incorporation of
the scalar-isoscalar and scalar-isovector sector into the meson-meson scattering
with the recently joint efforts in theory as well as experiment~\cite{Oller:1998hw,Guerrero:1998ei,Guo:2011pa,Kaminski:2009qg,Xiao:2001pt,Jia:2011zzd},
lattice study of $K \overline{K}$ scattering
is eventually evolving into a valuable enterprise.

Nevertheless, until now, no lattice study on  $K \overline{K}$ scattering
in the $I=0$ channel has been reported
chiefly because it is fairly arduous to rigorously calculate the rectangular $R_s$
and vacuum $V$ diagrams~\cite{Fu:2012ng}.
Motivated by lattice exploration on $K^+K^+$
scattering by NPLQCD~\cite{Beane:2007uh}
and encouraged by our preliminary lattice investigation on $K \overline{K}$ scattering
in the $I=1$ channel~\cite{Fu:2012ng},
we here further probe the $K \overline{K}$ scattering
in the $I=0$ channel.

It should be stressed that the physical evaluation of
the $K \overline{K}$ scattering length in the $I = 0$ channel is
a genuine two coupled-channel problem~\cite{Doring:2011vk,multi_channel},
which can be generally described
by $\pi\pi$ and $K\overline{K}$ channels
(we call $\pi\pi$ as channel $1$, while $K \overline{K}$ channel $2$).
The relevant $S$-matrix is a $2\times 2$ unitary matrix
comprising three real parameters~\cite{Doring:2011vk,multi_channel},
all of which are functions of the energy, and
a relationship among three parameters are provided
by the so-called generalized L\"uscher's formula~\cite{multi_channel}.
$K \overline{K}$ scattering in the $I = 1$ channel
should be handled analogously,
then lattice study of $\pi\eta$ scattering is extremely
wanted~\cite{Fu:2012ng,Doring:2011vk}.
In a nutshell, it is definitely  required to
involve  channel $1$ ($I=0$ $\pi\pi$ scattering)
for a physical evaluation of $K \overline{K}$ scattering length in the $I=0$ channel.
With our recent lattice attempts on channel $1$ in Ref.~\cite{Fu:2013ffa},
it is becoming more and more prospective to
solve this intriguing two coupled-channel problem,
and lattice study on channel $2$
is highly desirable.

In this paper, we make use of the MILC gauge configurations
with the $2+1$ flavors of the Asqtad improved
staggered sea quarks~\cite{Bernard:2010fr,Bazavov:2009bb}
to compute the channel $2$ ($I=0$ $K \overline{K}$ scattering).
The technique introduced in Ref.~\cite{Kuramashi:1993ka},
namely, the moving wall sources without gauge fixing,
is exploited to get the rigorous accuracy,
and a clear signal of attraction is observed.
As presented later, our lattice-calculated $K\overline{K}$
scattering is feasible and charming.

\section{Method}
\label{sec_method}
In keeping with the original derivations and notations in Ref.~\cite{Fu:2012ng},
we shortly review the required formulas for lattice study of
$K \overline{K}$ scattering enclosed in a torus.
To make this work self-sustaining,
all the basic pieces will be recapitulated succeedingly.

Given degenerate $u/d$ quark masses,
topologically only four quark line diagrams remain to contribute
to $K \overline{K}$ scattering amplitude in the $I=0$ channel,
which are exhibited in Fig.~\ref{fig:diagram_KK}.
\begin{center}
\includegraphics[width=8.0cm]{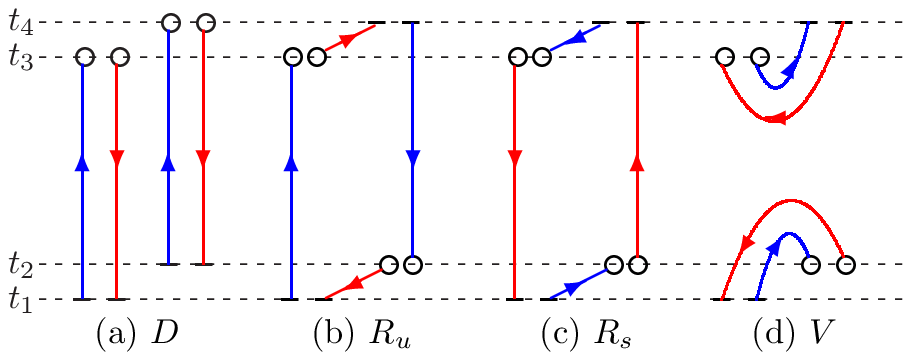}
\figcaption{ \label{fig:diagram_KK}
Quark-line diagrams contributing to the $I=0$ $K \overline{K}$
scattering amplitudes four-point functions.
Short bars render the wall sources.
Open circles are sinks for local $K/\overline{K}$ operators.
Blue and red lines indicate the light $u/d$ and strange $s$ quark lines, respectively.
}
\end{center}

To prohibit the convoluted color Fierz rearrangement of
quark lines~\cite{Kuramashi:1993ka},
we normally take $t_1 \ne t_2 \ne t_3 \ne t_4$
and  select $t_1 =0$, $t_2=1$, $t_3=t$, and $t_4 = t+1$, respectively.
In Fig.~\ref{fig:diagram_KK}, we illustrated the quark-line diagrams,
which are familiarly marked as direct ($D$), rectangular ($R_u$ and $R_s$),
and vacuum ($V$) diagrams, respectively~\cite{Fu:2012ng}.
It is not difficult to calculate the direct diagram.
Nonetheless, it is pretty challenging to compute other three
diagrams~\cite{Fu:2012ng}.

These diagrams were calculated by the moving wall sources technique
initiated by Kuramashi {\it et al.}~\cite{Kuramashi:1993ka},
i.e., each propagator, corresponding to a wall source
at $ t = 0, \cdots, T-1$, is represented by
$$
\sum_{x}{\cal D}_{n,x}G_t(x) = \sum_{\bf{x}}
\delta_{n,({\bf{x}},t)}, \quad 0 \leq t \leq T-1 ,
$$
where ${\cal D}$ is the Dirac quark matrix for quark action.
$D$, $R_u$, $R_s$ and $V$ are schematically displayed in Fig.~\ref{fig:diagram_KK},
and can be rewritten in terms of the quark propagators $G$~\cite{Fu:2012ng},
namely,
\begin{eqnarray}
\label{eq:dcr}
\hspace{-0.2cm}C_D(t_4,t_3,t_2,t_1)  \hspace{-0.3cm}&=&\hspace{-0.3cm}  \sum_{ {\bf{x}}_3,{\bf{x}}_4 }
\langle \mbox{Tr}
[G_{t_1}^{(u)\dag}({\bf{x}}_3, t_3) G^{(s)}_{t_1}({\bf{x}}_3, t_3)] \cr
&&\hspace{-0.3cm} \times
\mbox{Tr}[G_{t_2}^{(u)\dag}({\bf{x}}_4, t_4) G^{(s)}_{t_2}({\bf{x}}_4, t_4) ] \rangle,\cr
\hspace{-0.2cm}C_{R_u}(t_4,t_3,t_2,t_1) \hspace{-0.3cm}&=&\hspace{-0.3cm}
\sum_{ {\bf{x}}_2,{\bf{x}}_3 }
\langle \mbox{Tr}
[G_{t_1}^{(s)\dag}({\bf{x}}_2, t_2) G^{(u)}_{t_4}({\bf{x}}_2, t_2)  \cr
&&\hspace{-0.3cm} \times
G_{t_4}^{(s)\dag}({\bf{x}}_3, t_3) G^{(u)}_{t_1}({\bf{x}}_3, t_3) ] \rangle, \cr
\hspace{-0.2cm}C_{R_s}(t_4,t_3,t_2,t_1) \hspace{-0.3cm}&=&\hspace{-0.3cm}
\sum_{ {\bf{x}}_2, {\bf{x}}_3 }
\langle \mbox{Tr}
[G_{t_1}^{(u)\dag}({\bf{x}}_2, t_2) G^{(s)}_{t_4}({\bf{x}}_2, t_2)  \cr
&&\hspace{-0.3cm} \times
G_{t_4}^{(u)\dag}({\bf{x}}_3, t_3) G^{(s)}_{t_1}({\bf{x}}_3, t_3) ] \rangle, \cr
\hspace{-0.2cm}C_V(t_4,t_3,t_2,t_1) \hspace{-0.3cm}&=&\hspace{-0.3cm}
\sum_{{\bf{x}}_2,{\bf{x}}_3}
\Bigl\{ \langle \mbox{Tr}
[G_{t_1}^{(u) \dag}({\bf{x}}_2, t_2) G_{t_1}^{(s)}({\bf{x}}_2, t_2) ] \cr
&&\hspace{-0.3cm}  \times \mbox{Tr}
[G_{t_4}^{(u) \dag}({\bf{x}}_3, t_3) G_{t_4}^{(s)}({\bf{x}}_3, t_3) ] \rangle  -\cr
&&\hspace{-0.3cm} \langle \mbox{Tr}
[G_{t_1}^{(u) \dag}({\bf{x}}_2, t_2)
 G_{t_1}^{(s)}({\bf{x}}_2, t_2)] \rangle \cr
&&\hspace{-0.3cm} \times \langle \mbox{Tr}
[G_{t_4}^{(u) \dag} ({\bf{x}}_3, t_3) G_{t_4}^{(s)}({\bf{x}}_3, t_3) ]
\rangle \hspace{-0.1cm} \Bigl\},
\end{eqnarray}
where the Hermiticity properties of $G$ are employed
to nicely get rid of the $\gamma^5$ factors,
and the vacuum expectation value should be subtracted from the vacuum diagram.

The rectangular ($R_u$ and $R_s$) and vacuum ($V$) diagrams
unavoidably incur gauge-variant noise~\cite{Kuramashi:1993ka},
and which is usually suppressed through fulfilling the gauge field average
without gauge fixing~\cite{Kuramashi:1993ka,Fu:2011bz,Fu:2011wc,Fu:2012gf,Fu:2011xw,Fu:2012ng,Fu:2012tj}.
In the isospin limit, the $I=0$ $K\overline{K}$ four-point correlation function
can be stated on the basis of four diagrams~\cite{Fu:2012ng}:
$$
C_{K \overline{K}}(t) = D - N_f R_u - 2N_f R_s + 2V,
$$
where the extra factors $N_f$ in the valence fermion loops are addressed
by inserting the staggered-flavor factor $N_f$~\cite{Sharpe:1992pp}.
The four-fold degeneracy of the staggered sea
quarks is gotten rid of by performing the quadruplicate root of
the fermion determinant~\cite{Sharpe:1992pp}.

The energy $E_{K \overline{K}}$ of $K \overline{K}$ system is able to be gained
from the $K \overline{K}$ four-point function
at large $t$, whose asymptotic form behavior is provided as~\cite{Mihaly:1996ue}
\begin{eqnarray}
\label{eq:E_KaK}
\hspace{-0.6cm} C_{K \overline{K}}(t)  &=&
Z_{K \overline{K}}\cosh\left[E_{K \overline{K}}\left(t - \frac{T}{2}\right)\right] +\cr
&&\hspace{-1.0cm} (-1)^t Z_{K \overline{K}}^{\prime}
\cosh\left[E_{K \overline{K}}^{\prime} \left(t-\frac{T}{2}\right)\right] + \cdots,
\end{eqnarray}
where the oscillating term is
a typical characteristic of the Kogut-Susskind
formulation~\cite{Mihaly:1996ue}.

The magnitude of the center-of-mass scattering momentum $k$
is related to the energy $E_{K \overline{K}}$ of the $K \overline{K}$ system
in a torus of size $L$  by
\begin{equation}
\label{eq:MF_k_e}
k^2 = \frac{1}{4} \left( E_{K \overline{K}} \right)^2 - m_K^2,
\quad k = \frac{2\pi}{L} q ,
\end{equation}
where the dimensionless momentum $q \in \mathbb{R}$, and $m_K$ is kaon mass.

In practice, we also evaluate the ratios~\cite{Fu:2012ng}
\begin{equation}
\label{EQ:ratio}
R^X(t) =
\frac{ C_{K \overline{K}}^X(0,1,t,t + 1) }
     { C_K (0,t) C_K(1,t + 1) },
\quad  X=D, R_u, R_s \; {\rm and} \; V ,
\end{equation}
where $C_K(0,t)$ and $C_K(1,t+1)$ are kaon propagators with zero momentum.
It is worthwhile to stress that these ratios really assist
us to qualitatively comprehend the relevant physical quantities,
although we do not use them to estimate the energy shift.

It is worth mentioning that the contributions of the non-Nambu-Goldstone
kaons in the intermediate states is exponentially suppressed
for high $t$ due to their relatively larger masses
in contrast with those of Goldstone kaon~\cite{Sharpe:1992pp},
and this systematic error is overlooked in the present study.

\section{Simulation results}
In this work, we measured $K \overline{K}$ four-point function
on the $0.15$~fm MILC lattice ensemble of
$601$ $20^3 \times 48$ gauge configurations with
the $2+1$ flavors of the Asqtad-improved staggered sea quarks
with the bare quark masses
$am_{u/d}/am_s= 0.00484/0.0484$ and bare gauge coupling
$10/g^2 = 6.566$,
which has a physical volume around
$3.0$~fm~\cite{Bernard:2010fr,Bazavov:2009bb}.
The inverse lattice spacing $a^{-1}=1.373$~GeV~\cite{Bernard:2010fr,Bazavov:2009bb}.
The mass of the dynamical strange quark is
near to its physical value~\cite{Bernard:2010fr,Bazavov:2009bb},
and the light $u/d$ quark masses are equal.
In the present study, the valence strange quark mass is fixed to its
physical value~\cite{Bernard:2010fr,Bazavov:2009bb}.
It is noteworthy that the MILC gauge configurations are
generated using the fourth root of the staggered fermion
determinant, which nicely removes the quadruple degeneracy of
the staggered sea quarks~\cite{Bernard:2001av}.

The prevailing conjugate gradient approach is exploited to
achieve the requisite inverse matrix element of
the Dirac fermion matrix,
and the periodic boundary condition is enforced to
the spatial and temporal directions.
The propagators of $I=0$ $K \overline{K}$ scattering are
measured on each of the $48$ time slices,
and explicitly gathered at the end of the measurements.
For each kaon source, six fermion matrix inversions are demanded
for the possible three color selections.
Altogether, $288$ matrix inversions are carried out on a single configuration.
Consequently, the statistics are turned out to be significantly improved.

The individual amplitude ratios $R^X$ ($X=D$, $R_u$, $R_s$ and $V$)
of the $K \overline{K}$ four-point functions are
plotted in Fig.~\ref{fig:ratio} as the functions of $t$.
The direct amplitude $R^D$ is in the vicinity of unity,
indicating a weak interaction in this channel.
Following an initial increase up to $t \sim 4$,
the rectangular $R^{R_u}$ and $R^{R_s}$ amplitudes
exhibited a roughly linear decrease up until $t \sim 16$,
and the systematically oscillating behaviour in large $t$
is obviously noted~\cite{Mihaly:1996ue},
which is a typical feature of the staggered scheme~\cite{Fu:2012ng,Sharpe:1992pp,Mihaly:1996ue}.
It is interesting to notice that the rectangular  $R^{R_s}$ amplitude
oscillate much strongly than that of $R^{R_u}$.
The ``trumpet'' shape of the rectangular $R^{R_s}$ amplitude is
the signature for the lattice study of $I=0$ $K \overline{K}$ scattering
with staggered scheme,
which is turned out to the most important parts of $I=0$ $K \overline{K}$ scattering.
The intensive oscillation of the rectangular
$R^{R_s}$ amplitude is due to the staggered scheme,
nevertheless, its quantitative mass dependence
is not yet comprehensive to us and demands an additional investigations.

\begin{center}
\includegraphics[width=8.0cm]{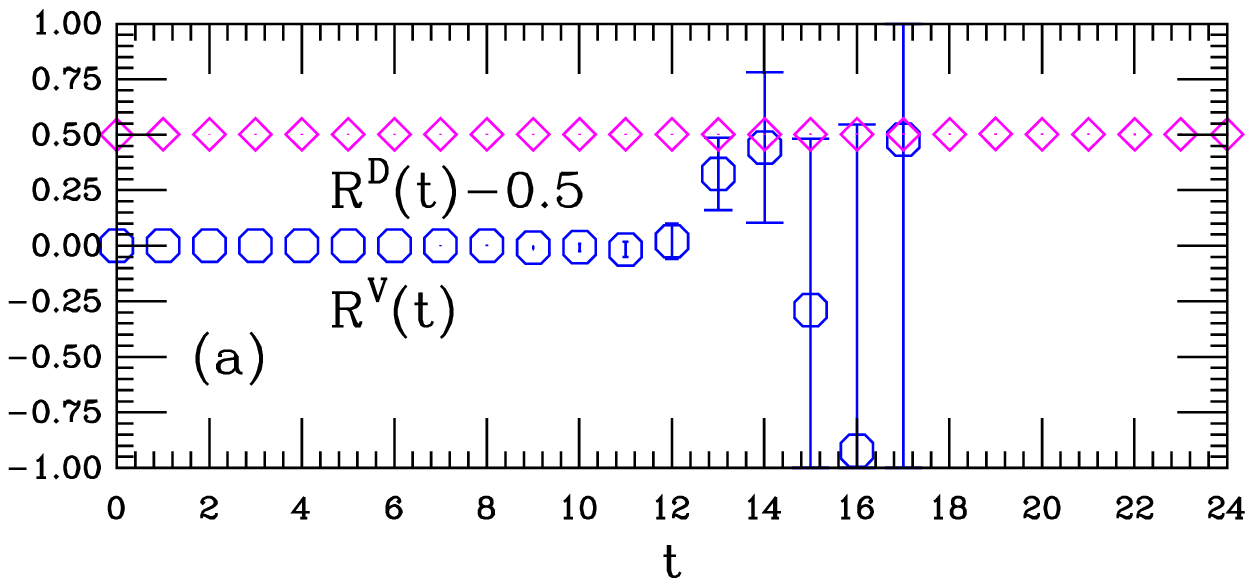}
\includegraphics[width=8.0cm]{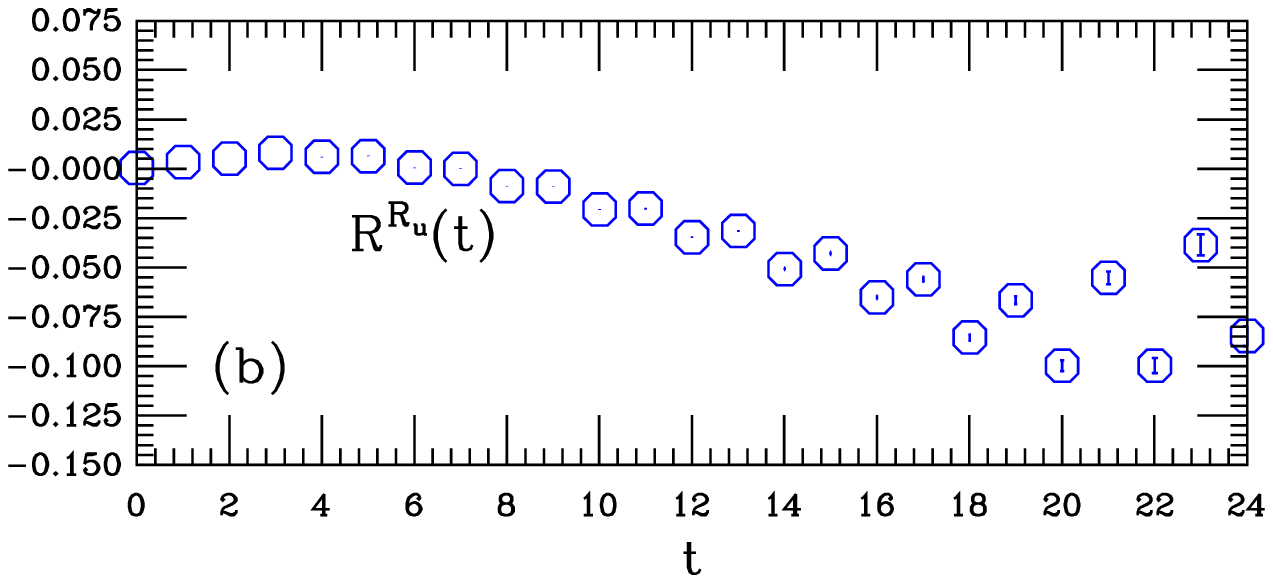}
\includegraphics[width=8.0cm]{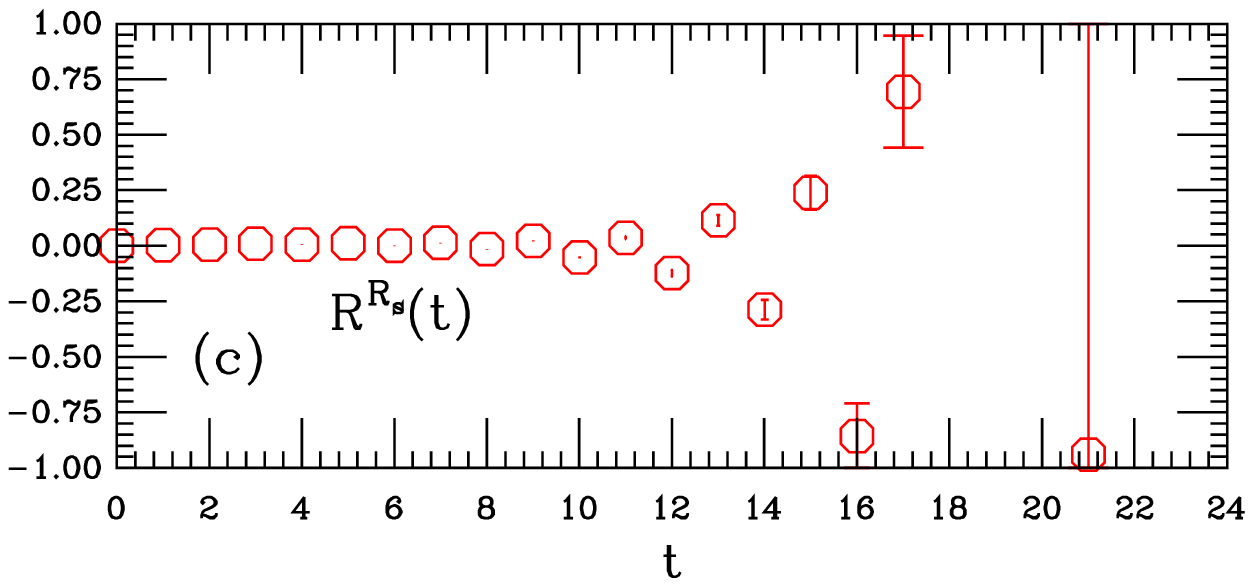}
\figcaption{
Separate amplitude ratios $R^X$ ($X=D$, $R_u$, $R_s$ and $V$)
of $K \overline{K}$ four-point functions as the functions of $t$.
(a) Direct diagram shifted by $0.5$ (magenta diamonds) and vacuum diagram (blue octagons);
(b) Rectangular diagram $R_u$;
(c) Rectangular diagram $R_s$.
\label{fig:ratio}
}
\end{center}

The vacuum amplitude is fairly small up to $t \sim 8$,
and quickly lost signals after that.
On the basis of the Lepage's analytical arguments~\cite{Lepage:1989hd},
we are able to deduce that the error of the disconnected diagram
is generally independent of $t$ and grows exponentially
as $\displaystyle e^{2m_K t}$~\cite{Fu:2012ng},
meanwhile the errors of the rectangular $R_u$ and $R_s$ diagrams
grow exponentially as $\displaystyle e^{(2m_K - m_{s\bar{s}})t}$ and
$\displaystyle e^{(2m_K - m_{\pi})t}$, respectively,
where $m_{s\bar{s}}$ is mass of a fictitious flavor nonsinglet
meson $s\bar{s}$~\cite{Aubin:2004wf}
(see Ref.~\cite{Bernard:2001av} for the evaluation of $m_{s\bar{s}}$)~\footnote{
For the MILC gauge configuration used in Ref.~\cite{Fu:2012ng},
it is ready to verify that $m_{\pi} \approx 2m_K - m_{s\bar{s}} $
and $m_{s\bar{s}} \approx 2m_K - m_{\pi}$.
}.
This means that the calculation of the rectangular $R_u$ diagram
is relatively cheap, while the calculations of
the rectangular $R_s$ and vacuum $V$ diagrams
are fairly challenging.

The magnitudes of the errors of these ratios are in quantitative agreement with
these predictions as shown in Fig.~\ref{fig:error}.
Fitting these errors $\delta R^X(t)$ by a single exponential fit ansatz
$\displaystyle \delta R^X(t) \sim  e^{\mu_X t}$ over $8 \le t \le 18$,
we secure $\displaystyle a \mu_X = 0.2457$, $0.5813$ and $0.7575$
for $X=R_u$, $R_s$ and $V$, respectively,
which can be fairly compared with our measurements of
$a m_\pi = 0.1750(1)$, $a m_K = 0.3780(1)$ and
$a m_{s\bar{s}} = 0.5012(2)$ on the same MILC gauge configurations in this work.
In principle, the vacuum amplitude can be reasonably granted
to be still small for large $t$, and discarded in the remaining analysis~\cite{Kuramashi:1993ka}.
Nevertheless, we still incorporate it such that this calculation has integrity.

\begin{center}
\includegraphics[width=8.0cm]{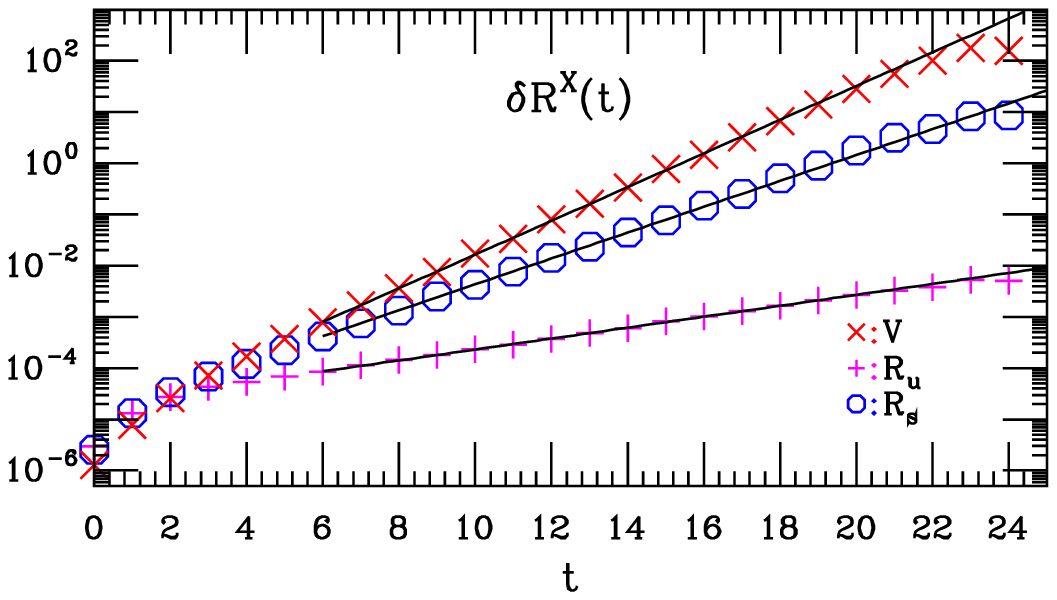}
\figcaption{\label{fig:error}
Errors of the ratios $R^X(t)$ ($X=R_u$, $R_s$ and $V$) as the functions of $t$.
Solid lines are single exponential fits over $8 \le t \le 16$.
}
\end{center}

The point operator has a big overlap with excited states,
and it is traditional to use the wall-source operator together with a point
sink to efficiently reduce these overlap~\cite{Aubin:2004fs}.
In practice, we calculate the kaon propagators by
$$
C_K^{\rm WP}(t) =
\langle 0|K^\dag (t) W_{K^+}(0) |0\rangle,
$$
where $K^+=\bar{s}\gamma_5 u$, $W_{K^+}$ is wall-source operator for $K^+$ meson~\cite{Bernard:2001av,Aubin:2004wf}.
To ease the notation,
the summation over the spatial coordinates of sink is not written out,
and we adopt the shorthand notation:
``WP'' for the wall-source point-sink propagators~\cite{Aubin:2004wf,Fu:2013ffa}.
In practice, we exploited the available wall source quark propagators $G$,
which were used to calculate $\pi\pi$ amplitudes,
to evaluate kaon propagators,
\begin{equation}
\hspace{-0.3cm}C_K^{\rm WP}(t) = \sum_{ {\bf{x}}_2 }
\langle \mbox{Tr}
G_{t_1}^{(u)\dag}({\bf{x}}_2, t_2) G^{(s)}_{t_1}({\bf{x}}_2, t_2) \rangle,
\end{equation}
where $t=t_2-t_1$, and we also evaluated kaon propagators
on each of the $48$ time slices,
and explicitly gathered these results, and
the statistics are proved to be remarkably improved.

Ignoring the excited state contaminations, the kaon mass $m_K$
can be obtained at large $t$ with a single exponential
fit ansatz~\cite{Bazavov:2009bb}
\begin{eqnarray}
\label{eq:pi_fit_WP}
C_K^{\rm WP}(t) &=& A_K^{\rm WP} \left[e^{-m_K t}+e^{-m_K(T-t)}\right],
\end{eqnarray}
where $A_K^{\rm WP}$ is overlapping amplitude,
which is subsequently exploited to estimate the wraparound
contributions~\cite{Gupta:1993rn,Umeda:2007hy}, and
for easier notation the superscript WP in $A_{K}$ is leaved out
in the rest of analyses.

From Fig.~\ref{fig:ratio}, we clearly observe
a contamination from the wraparound
effects~\cite{Gupta:1993rn,Umeda:2007hy}
approximately starting at $t = 16 \sim 18$,
which acts as a constant contribution and
spoils $K \overline{K}$ four-point functions in large $t$
(in particular around $T/2$),
since one of two kaons can propagate $T-t$ time steps backwards
with the periodic boundary condition imposed in the temporal direction.
Luckily, it is approximately suppressed by the factor
$
\exp \left( -  m_K  T \right) / \exp \left( - 2m_K t \right)
$
as compared with the forward propagation of the $K \overline{K}$ state~\cite{Gupta:1993rn,Umeda:2007hy}.
In practice, we can carefully choose the fitting ranges
obeying $t_{\rm max} \ll T/2$ to reduce this unwanted pollution~\cite{Gupta:1993rn}.

Additionally, if kaon mass is small enough,
this wraparound term is turned out to be involved for an acceptable fit,
\begin{eqnarray}
\label{eq:E_K_K}
\hspace{-0.4cm} C_{K \overline{K}}(t)  \hspace{-0.2cm}&=&\hspace{-0.2cm} C+
Z_{K \overline{K}}\cosh\left[E_{K \overline{K}}\left(t - \tfrac{T}{2}\right)\right]
+\cr
&&\hspace{-0.2cm}(-1)^t Z_{K \overline{K}}^{\prime}
\cosh\left[E_{K \overline{K}}^{\prime} \left(t-\tfrac{T}{2}\right)\right] + \cdots,
\end{eqnarray}
where $C$ is a constant relevant to the wraparound pollution,
which can be calculated as well though~\cite{Gupta:1993rn,Umeda:2007hy}
\begin{equation}
\label{eq:C_fake_diagram}
C = 2A_K^2 e^{-m_K T}.
\end{equation}

Using Eq.~(\ref{eq:pi_fit_WP}), the overlapping amplitudes
$A_K$ and kaon masses $m_K$ are accurately secured
from corresponding kaon correlators
and these values are fairly accurate
to evaluate the wraparound terms by Eq.~(\ref{eq:C_fake_diagram}).
All of these values are listed in Table~\ref{tab:C_C}.
\begin{center}	
\tabcaption{\label{tab:C_C}
Summaries of evaluated wraparound term
from  overlapping amplitude $A_K$ and kaon mass $am_K$.
The first and second block  shows our fitted $A_K$
and $am_K$,
and column $3$ gives  the wraparound  term estimated
by Eq.~(\ref{eq:C_fake_diagram}),
where its errors are combined from
the errors of $A_K$ and $m_K$.
}
\footnotesize
\begin{tabular*}{80mm}{c@{\extracolsep{\fill}}ccc}
\toprule
$m_K$  & $A_K $ & $C$    \\
\hline
$0.37803(14)$ & $558.85(90)$ & $0.00822(6)$  \\
\bottomrule	
\end{tabular*}
\end{center}	

In practice, $K \overline{K}$ four-point correlation function is fit
by denoting a fitting range $\rm {[D_{min},D_{max}]}$ and
varying the values of the minimum fitting distances $\rm D_{min}$
and with the maximum fitting distance $\rm D_{max}$ either at $T/2$
or where the fractional errors  transcended about $30\%$
for two successive time slices.
In addition, the fitting parameter $C$
was firmly confined by priors to the lattice-estimated
wraparound term $C$  recorded in Table~\ref{tab:C_C}.
The ``effective energy'' plots as a function
of $\rm D_{min}$ are displayed in Fig.~\ref{fig:plateau_KK_I0}.
The central value and statistical error per time slice were
estimated by the standard jackknife procedure.
\begin{center}
\includegraphics[width=8.0cm]{KK_I0_00484.eps}
\figcaption{ \label{fig:plateau_KK_I0}
Effective energy plots as the function of
$\rm D_{min}$ for $K \overline{K}$ scattering in the $I=0$ channel in lattice units.
The magenta diamond points indicate the results omitting vacuum diagram (Vout),
while the blue square points show the results involving vacuum diagram (Vin).
}
\end{center}

In this work, the energies $E_{K \overline{K}}$ of the $I=0$ $K \overline{K}$ system
were extracted from the effective energy plots, and
we pick up the energy $E_{K \overline{K}}$ by the full consideration of
a plateau in the energy as a function of $\rm D_{min}$,
a good confidence level ($\chi^2/{\rm dof}$),
and enough large $\rm D_{min}$  to suppress the excited states~\cite{Fu:2012ng}.
From Fig.~\ref{fig:plateau_KK_I0}, we keenly observe
the plateau in the effective energy is pretty short.
This is not surprise for us since, as we explained above,
the signals of the rectangular $R_s$ and $V$ diagrams are fairly noisy,
as a consequence, we should select small $\rm D_{min}$
to get an acceptable fits.
On the other hand, if we choose pretty small $\rm D_{min}$,
then the pollution from excited states cannot be neglected.
In practice, only the time range $6\le D_{\rm min} \le 7$ can be considered.
This indicates that the robust calculations of the rectangular $R_s$ and $V$ diagrams
are high desirable for the more sophisticated
lattice computation of the $I=0$ $K \overline{K}$ scattering in the future.

The $K \overline{K}$ four-point functions
in the $I=0$ channel are exhibited in Fig.~\ref{fig:KpKm},
where the fitted functional forms are compared with lattice data.
The systematically oscillating features
are expectedly observed in large $t$~\cite{Mihaly:1996ue}.
The scattering momentum $k^2$ in lattice units calculated by Eq.~(\ref{eq:MF_k_e}),
along with the fitted $E_{K \overline{K}}$,
its fitting range and fitting quality ($\chi^2/{\rm dof}$)
are listed in Table~\ref{tab:fitting_results_KpKm}.
The statistical errors of the scattering momentum $k^2$ are combined from
the uncertainties of the energy $E_{K \overline{K}}$ and kaon mass $m_K$ in quadrature.

\begin{center}
\includegraphics[width=8.0cm]{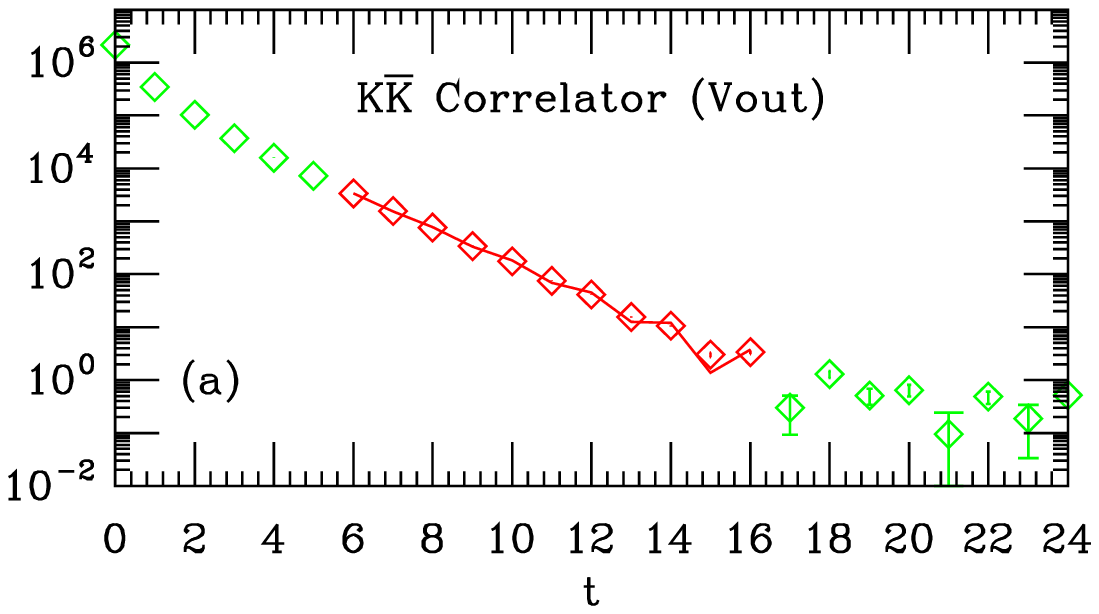}
\includegraphics[width=8.0cm]{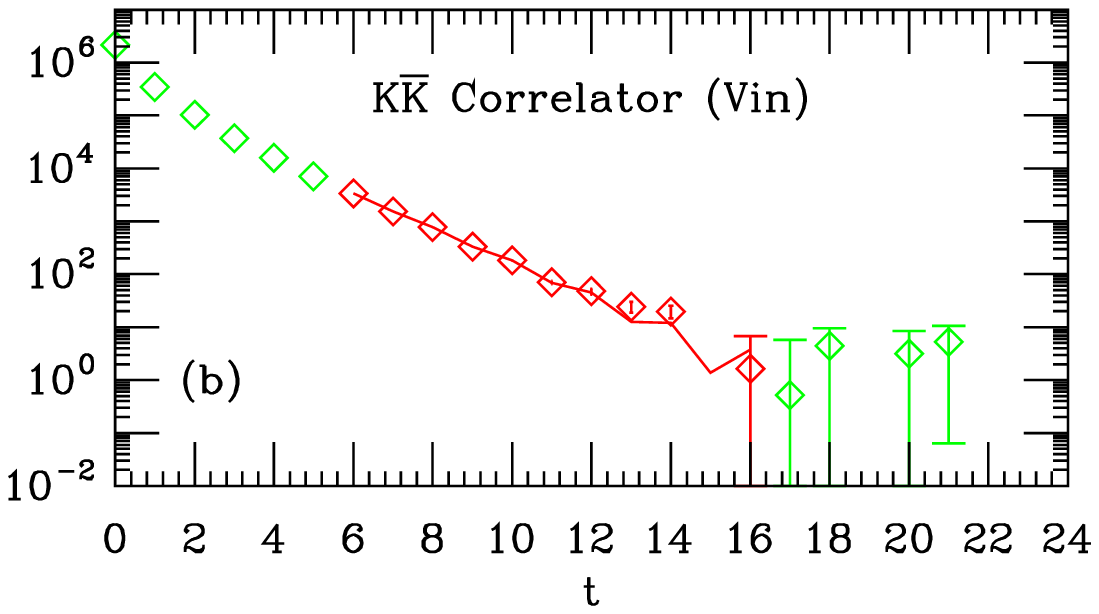}
\figcaption{\label{fig:KpKm}
$K \overline{K}$ four-point function.
Fitting range is specified by points and fitted
lines in red.
(a) The results omitting vacuum diagram, and
(b) The results involving vacuum diagram.
Occasional points with negative mean values are not illustrated.
}
\end{center}

\begin{center}	
\tabcaption{\label{tab:fitting_results_KpKm}
The energy $E_{K \overline{K}}$ and scattering momentum $k^2$ in lattice units.
Column $2$ gives the results omitting vacuum diagram.
Column $3$ shows those incorporating vacuum diagram.
Fitting range and fit quality  are also provided.
}
\footnotesize
\begin{tabular*}{80mm}{@{\extracolsep{\fill}}lcc}
\toprule
&{\rm Fit A (Vout)}             &{\rm Fit B (Vin)}               \\
\hline
$E_{K \overline{K}}$              &$0.74764(98)$  & $0.7481(31)$  \\
$k^2$                             &$-0.00597(72)$ & $-0.00568(221)$ \\
{\rm Range [$\rm D_{min},D_{max}$]}    &$6-16$         & $6-16$          \\
$\chi^2/{\rm dof}$                &$12.5/6$       & $8.5/6$         \\
\bottomrule	
\end{tabular*}
\end{center}

From Table~\ref{tab:fitting_results_KpKm},
we clearly note that the interaction for the $I=0$ $K\overline{K}$ scattering
is attractive,
which is in good agreement with the theoretical (or phenomenological) predictions~\cite{Oller:1998hw,Guerrero:1998ei,Guo:2011pa,Kaminski:2009qg,Xiao:2001pt,Jia:2011zzd}.
We are particularly satisfactory about these results
considering the aforementioned extremely difficulties of the lattice calculations
for the rectangular $R_s$ and vacuum $V$ diagrams~\cite{Fu:2012ng}.
Nonetheless, we are clearly aware that
there still have a large statistical error for Fit A (omitting the vacuum term: Vout)
and rather big error in Fit B (involving the vacuum diagram: Vin)~\cite{Fu:2012ng},
which are absolutely needed to be improved in the future.

\section{Conclusions and outlooks}
In this initiatory work, we carried out a lattice calculation of
$K \overline{K}$ scattering in the $I=0$ channel, where both rectangular
$R_s$ diagram and vacuum $V$ diagram play a crucial role.
We took advantage of the moving wall source without
gauge fixing~\cite{Kuramashi:1993ka}
to calculate all the four diagrams classified in Ref.~\cite{Fu:2012ng},
and clearly observed an attractive signal in this channel
with the consideration of wraparound effect,
which is in well agreement with the theoretical (or phenomenological) predictions~\cite{Oller:1998hw,Guerrero:1998ei,Guo:2011pa,Kaminski:2009qg,Xiao:2001pt,Jia:2011zzd}.

Confronting the afore-discussed intrinsic difficulties
of lattice calculations on the rectangular $R_s$ and vacuum $V$ diagrams,
we still launched a lattice study to
examine $I=0$  $K\overline{K}$  scattering
with scant computational resources.
Unquestionably, due to low statistics, the uncertainties are rather large,
so these simulations are definitely in the preliminary stage,
and various sources of systematic error need to be elucidated comprehensively.
In principle, the signal-to-noise ratio is able to be further improved
by launching the same computation on the lattice ensembles with a smaller kaon mass.
Another important way to improve the statistics is
to use more lattice gauge configurations.
Moreover, according to the aforementioned discussions,
the error of the disconnected $V$ and rectangular $R_s$ diagrams
grow exponentially as $\displaystyle e^{2m_K t}$ and
$\displaystyle e^{(2m_K - m_{\pi})t}$, respectively,
where pion mass $m_\pi$ and kaon mass $m_K$
are measured in lattice units ($a m_\pi$ and $a m_K$).
Since for the same pion and kaon masses,
the extracted values of $a m_\pi$ and $a m_K$ are smaller
for the simulations on a larger lattice volume,
consequently, performing the calculation on a larger volume with a smaller kaon mass
maybe will be the most feasible way to improve the statistics.

Since it is a pioneer trial of lattice examination of
the $I=0$ $K \overline{K}$ scattering,
our principal aim is to demonstrate this approach in a conceptually clean way,
and achieve out our anticipated objective.
Most of all, we show that the moving wall source techniques can be employed
to investigate $K \overline{K}$ scattering in the $I=0$ channel.
This raises a good prospect that this strategy maybe be successfully exploited
to deal with two coupled-channel problem~\cite{Doring:2011vk,multi_channel}.
With our lattice efforts on channel $1$ in Ref.~\cite{Fu:2013ffa}, at present,
if we are able to compute the $\pi\pi \to K\overline{K}$ scattering in the
$I=0$ channel, in principle, we can solve this problem.
We unceasingly appeal for enough computational resources to accomplish
this peculiar and challenging task,
and preparing a series of lattice studies toward these targets.

\section*{Acknowledgments}
The author owes a great deal to the MILC Collaboration
for using the Asqtad lattice ensembles and MILC codes.
We sincerely thanks Carleton DeTar for providing us the fitting software.
The author feel much indebted Paul Kienzle for teaching us
computation skills for this work during my work in NIST, Gaithersburg, USA.
We thankfully acknowledges the computational resources
and technical supports provided by Institute of
Nuclear Science and Technology, Sichuan University
(specially Hou Qing, Zhu An,  Ning Liu, Jun Wang and  Mantian Liu).
Numerical computations for this work were
carried out at HP, AMAX, and CENTOS workstations.

\end{multicols}

\vspace{-1mm}
\centerline{\rule{80mm}{0.1pt}}
\begin{multicols}{2}

\end{multicols}

\clearpage

\end{CJK*}

\begin{thebibliography}{10}
\bibitem{Fu:2012ng} Z. Fu
Eur. Phys. J. {\bf C 72} (2012) 2159.

\bibitem{Doring:2011vk} M.~Doring, U.~-G.~Meissner, E.~Oset and A.~Rusetsky,
Eur.\ Phys.\ J.\ {\bf A 47} (2011) 139.

\bibitem{Winter:2006vd} P. Winter {\it et al.},
Phys. Lett. {\bf B 635} (2006) 23.

\bibitem{Dzyuba:2008fi} A. Dzyuba  {\it et al.},
Phys. Lett. {\bf B 668} (2008) 315.

\bibitem{Maeda:2008mx} Y. Maeda {\it et al.},
Phys. Rev.  {\bf C 79} (2009) 018201.

\bibitem{Xie:2010md} J. J. Xie  and C. Wilkin,
Phys. Rev. {\bf C 82} (2010) 025210.

\bibitem{Silarski:2009yt} M. Silarski {\it et al.},
Phys. Rev. {\bf C 80} (2009) 045202.

\bibitem{Silarski:2010ae} M.~Silarski,
Int. J. Mod. Phys. {\bf A 26} (2011) 539.

\bibitem{Li:1997zb} G.~-Q. Li,  C. H. Lee  and G. E. Brown,
Nucl. Phys. {\bf A 625} (1997) 372.

\bibitem{Detmold:2008yn} W. Detmold, K. Orginos, M. J. Savage and A. Walker-Loud,
Phys. Rev. {\bf D 78} (2008) 054514.

\bibitem{Oller:1998hw} Oller J. A., Oset E. and Pelaez J. R.,
Phys. Rev. {\bf D 59} (1999) 074001.

\bibitem{Guerrero:1998ei} F. Guerrero and J. A. Oller,
Nucl. Phys. {\bf B 537} (1999) 459.

\bibitem{Guo:2011pa} Z. H. Guo and J. A. Oller,
Phys. Rev. {\bf D 84} (2011) 034005.

\bibitem{Kaminski:2009qg} R. Kaminski, G. Mennessier and S. Narison,
Phys. Lett. {\bf B 680} (2009) 148.

\bibitem{Xiao:2001pt} Z.~Xiao and H.~-q.~Zheng,
Commun.\ Theor.\ Phys.\  {\bf 48} (2007) 685.

\bibitem{Jia:2011zzd} E.~-W.~Jia and H.~-R.~Pang,
Chin.\ Phys.\ Lett.\  {\bf 28} (2011) 061401.

\bibitem{Beane:2007uh} Beane S. R.,  Luu T. C., Orginos K., Parrene A., Savage M. J.,
Torok A., and Walker-Loud A.,
Phys. Rev. {\bf D 77} (2008) 094507.

\bibitem{multi_channel}
S.~He, X.~Feng and C.~Liu,
JHEP {\bf 0507} (2005) 011;
M.~Doring and U.~G.~Meissner,
JHEP {\bf 1201} (2012) 009;
M.~T.~Hansen and S.~R.~Sharpe,
Phys.\ Rev.\ {\bf D 86} (2012) 016007;
C.~Liu, X.~Feng and S.~He,
Int.\ J.\ Mod.\ Phys.\  {\bf A 21} (2006) 847;
N.~Li and C.~Liu,
Phys.\ Rev.\ {\bf D 87} (2013) 014502.

\bibitem{Fu:2013ffa} Z.~Fu,
Phys.\ Rev.\ {\bf D 87} (2013) 074501.

\bibitem{Bernard:2010fr} C. Bernard {\it et al.},
Phys. Rev. {\bf D 83} (2011) 034503 .


\bibitem{Bazavov:2009bb} A. Bazavov {\it et al.},
Rev. Mod. Phys. {\bf 82} (2010) 1349 .

\bibitem{Kuramashi:1993ka} Y. Kuramashi, M. Fukugita,  H. Mino,  M. Okawa and A. Ukawa,
Phys. Rev. Lett. {\bf 71} (1993) 2387;
M. Fukugita, Y. Kuramashi, M. Okawa, H. Mino and A. Ukawa,
Phys. Rev. {\bf D 52} (1994) 3003.

\bibitem{Fu:2011bz} Z. Fu,
Commun. Theor. Phys. {\bf 57} (2012) 78.

\bibitem{Fu:2011wc} Z. Fu,
Phys. Rev. {\bf D 85} (2012) 074501.

\bibitem{Fu:2011xw} Z. Fu,
J. High Energy Phys {\bf 01} (2012) 017.

\bibitem{Fu:2012gf} Z. Fu,
J. High Energy Phys {\bf 07} (2012) 142.

\bibitem{Fu:2012tj} Z.~Fu and K.~Fu,
Phys.\ Rev.\ {\bf D 86} (2012) 094507.

\bibitem{Sharpe:1992pp} Sharpe S. R., Gupta R. and Kilcup G. W.,
Nucl. Phys.  {\bf B383} (1992) 309.

\bibitem{Mihaly:1996ue} Mihaly A., Fiebig H.R., Markum H. and Rabitsch K.,
Phys. Rev. {\bf D 55} (1997) 3077.

\bibitem{Bernard:2001av} C. W. Bernard, T. Burch, K. Orginos, D. Toussaint, T. A. DeGrand,
C. E. Detar, S. Datta, S. A. Gottlieb, U. M. Heller and R. Sugar,
Phys. Rev. {\bf D 64} (2001) 054506.

\bibitem{Lepage:1989hd} G. P. Lepage,
CLNS-89-971, 1990.

\bibitem{Aubin:2004wf} C.~Aubin, C.~Bernard, C.~DeTar, J.~Osborn, S.~Gottlieb, E.~B.~Gregory,
D.~Toussaint and U.~M.~Heller {\it et al.},
Phys.\ Rev.\ {\bf D 70} (2004)  094505.

\bibitem{Aubin:2004fs} C.~Aubin {\it et al.}  [MILC Collaboration],
Phys.\ Rev.\ {\bf D 70} (2004) 114501.

\bibitem{Gupta:1993rn} R.~Gupta, A.~Patel and S.~R.~Sharpe,
Phys.\ Rev.\ {\bf D 48} (1993) 388.

\bibitem{Umeda:2007hy} T.~Umeda,
Phys.\ Rev.\ {\bf D 75} (2007) 094502.



\end{thebibliography}
\end{document}